# Tokamak elongation – how much is too much? II Numerical results

J. P. Lee[1,2], A. Cerfon[1], J. P. Freidberg[2]

**Abstract**

The analytic theory presented in Paper I is converted into a form convenient for numerical analysis. A fast and accurate code has been written using this numerical formulation. The results are presented by first defining a reference set of physical parameters based on experimental data from high performance discharges. Numerically obtained scaling relations of maximum achievable elongation ($\kappa_{\max}$) versus inverse aspect ratio ($\varepsilon$) are obtained for various values of poloidal beta ($\beta_p$), wall radius ($b/a$), and feedback capability parameter ($\gamma\tau_w$) in ranges near the reference values. It is also shown that each value of $\kappa_{\max}$ occurs at a corresponding value of optimized triangularity ($\delta$), whose scaling is also determined as a function of $\varepsilon$. The results show that the theoretical predictions of $\kappa_{\max}$ are slightly higher than experimental observations for high performance discharges as measured by high average pressure. The theoretical $\delta$ values are noticeably lower. We suggest that the explanation is associated with the observation that high performance involves not only MHD considerations, but also transport as characterized by $\tau_E$. Operation away from the MHD optimum may still lead to higher performance if there are more than compensatory gains in $\tau_E$. Unfortunately, while the empirical scaling of $\tau_E$ with the elongation ($\kappa$) has been determined, the dependence on $\delta$ has still not been quantified. This information is needed in order to perform more accurate overall optimizations in future experimental designs.

1. Courant Institute of Mathematical Sciences, NYU, New York City NY
2. Plasma Science and Fusion Center, MIT, Cambridge MA

## 1. Introduction

In Paper II we convert the analytic formulation of the variational principle derived in Paper I (Freidberg et al. 2015) into a form suitable for numerical analysis. A code has been written based on this analysis that allows us to quickly and accurately calculate the dependence of elongation $\kappa$ and triangularity $\delta$ on inverse aspect ratio $\varepsilon$ for various values of poloidal beta $\beta_p$, wall radius $b / a$, and feedback parameter $\gamma\tau_w$. These scaling dependencies provide useful information for the optimization of plasma shape against asymmetric $n = 0$ MHD instabilities, which are the cause of vertical disruptions.

For perspective it is worth noting that there have been many numerical investigations of $n = 0$ MHD stability for a plasma surrounded by a perfectly conducting wall (Laval and Pellat 1973; Wesson and Skyes 1975; Becker and Lackner 1977) or with a resistive wall (Wesson 1975; Wesson 1978; Lazarus et al. 1991). In these studies, the growth rate of the mode is obtained either by directly solving the equations of motion or by minimizing $\delta W$. These studies have provided valuable insight into the vertical stability of a tokamak including design guidelines for optimizing performance. However, they have not focused on including the effect of feedback on the scaling of maximum elongation with aspect ratio which is the main goal of the present paper.

In comparison to previous studies our results are obtained using a somewhat more realistic model of the wall geometry. On the other hand our results are somewhat more restrictive in that we use only the well-known Solov'ev profile for the equilibrium (Solov'ev 1968). The Solov'ev profile provides accurate scaling with respect to plasma pressure and shape but is limited in its ability to take into account the effect of current profile on stability; that is the internal inductance per unit length is always on the order of $l_i \sim 0.4$ for all of our results. Still, the general scaling relations are accurate (see for instance Bernard et al. 1978) and, importantly, the profile leads to significant savings in computer time. The savings result from the fact that the Green's theorem for the solution of the vacuum region can also be utilized in the plasma region thereby reducing the 2-D stability problem into a 1-D problem.

An outline of the analysis is as follows. The numerical formulation of the variational principle is based on a combination of Fourier analysis and the application of Green's theorem. The analysis is carried out in terms of the perturbed magnetic flux. A substantial simplification occurs for the Solov'ev profiles because the perturbed poloidal magnetic field in the plasma turns out to be a vacuum field; that is, the perturbed toroidal current is zero. In this case the standard volume integral for the plasma energy $\delta W_F$ can be converted to a simple surface integral, thus transforming the 2-D problem into a 1-D problem. This is not true for more general profiles.

The basic strategy is to introduce Fourier expansions for the flux and its normal derivative on two surfaces, the plasma and wall. The corresponding Fourier amplitudes are the unknowns in the problem. Furthermore, the normal derivative amplitudes are related to the flux amplitudes through the solution of the vacuum flux equation, (i.e. $\Delta^*\psi = 0$), a step that is conveniently carried out using Green's theorem.

The end result is a classic minimizing principle that consists of the ratio of quadratic terms in the Fourier amplitudes subject to a series of linear constraints arising from the application of Green's theorem. Also, the matrix elements contain the resistive wall feedback parameter $\gamma\tau_w$, which appears in a simple linear form. The calculation thus reduces to a standard linear algebra problem in which, after some analysis, all the matrices are shown to be real and symmetric.

A summary of our results with respect to the effect of feedback on vertical stability is as follows. For values of $\gamma\tau_w$ similar to present day high performance tokamaks we find that the addition of feedback substantially increases the achievable elongation, typically from about 1.17 to 2.06 at $\varepsilon \approx 0.3$. Equally important we show that the achievable value of $\kappa$ decreases as $\varepsilon$ gets smaller for any value of $\gamma\tau_w$. In addition we find that at each value of maximum elongation ($\kappa_{\max}$) there is a corresponding value of optimized triangularity ($\delta$) whose scaling is also determined as a function of $\varepsilon$. Theoretical predictions of $\kappa_{\max}$ are slightly higher than experimental observations for high performance discharges as measured by high average pressure. Theoretical $\delta$ values are noticeably lower. The explanation is likely associated with the fact that high performance involves not only MHD considerations, but also transport as characterized by $\tau_E$. Operation away from the MHD optimum may still lead to higher performance if there are more than compensatory gains in $\tau_E$. Unfortunately, while the empirical scaling of $\tau_E$ with $\kappa$ is known, the dependence on $\delta$ has yet to be quantified. This information is needed in order to perform more accurate overall optimizations in future experimental designs.

The presentation of the analysis and results begins with Section 2, where we convert the Lagrangian integral derived in Paper I into a set of surface integrals by making use of the Solov'ev profile. In Section 3, the surface integrals are simplified by expressing them in the form of a symmetric matrix $\mathbf{W}$ and a vector variable of poloidal Fourier mode amplitudes of the perturbed fluxes and their normal derivatives. In Section 4, the constraints between the perturbed fluxes and their normal derivatives are obtained by utilizing the well-known Green's function for a vacuum region. In Section 5, we describe how numerical solutions are obtained by iterating the plasma parameters in order to make the minimum eigenvalue of $\mathbf{W}$ in the subspace of the constraints equal to zero. The eigenvalues are efficiently calculated using a QR decomposition. In Section 6, the parameter space of the numerical calculations is chosen by introducing (1) a reasonably realistic wall geometry model, and (2) a reference case of numerical input parameters determined by examining high performance experimental discharges from several tokamaks. Finally, the numerical results and discussion are given in Section 7 and Section 8, respectively.

## 2. The starting point

The starting point for the analysis is the Lagrangian integral for the variational principle repeated here for convenience,

$$
\begin{aligned}
&L = \delta W_F + \delta W_{V_I} + \delta W_{V_0} + \alpha W_D \\
&\delta W_F = \frac{1}{2\mu_0}\int_{V_P}\left[\frac{(\nabla\psi)^2}{R^2} - \left(\mu_0 p'' + \frac{1}{2R^2}F^{2\prime\prime}\right)\psi^2\right]d\mathbf{r} + \frac{1}{2\mu_0}\int_{S_P}\left(\frac{\mu_0 J_\phi}{R^2 B_p}\psi^2\right)dS \\
&\delta W_{V_I} = \frac{1}{2\mu_0}\int_{V_I}\frac{(\nabla\hat{\psi})^2}{R^2}d\mathbf{r} \\
&\delta W_{V_O} = \frac{1}{2\mu_0}\int_{V_O}\frac{(\nabla\hat{\hat{\psi}})^2}{R^2}d\mathbf{r} \\
&W_D = \frac{1}{2\mu_0}\int_{S_W}\frac{\hat{\psi}^2}{R^2}dS
\end{aligned}
\tag{1}
$$

where $\alpha = \gamma\mu_0\sigma d$ , with $\gamma$ the growth rate of the vertical instability, $\sigma$ the wall conductivity, and $d$ the thickness of the (thin) wall (see paper I). Note that in order to avoid the presence of multiple indices on $\psi$ later in the article, we have slightly modified the notation used in Paper I by deleting subscripts on the perturbed flux. Instead, hereafter $\psi$ is the flux in the plasma, $\hat{\psi}$ is the flux in the inner vacuum region, and $\hat{\hat{\psi}}$ is the flux in the outer vacuum region. At this point, it is interesting to observe that for the special case of Solov'ev profiles, $p'' = F^{2\prime\prime} = 0$, showing that the contribution from for the plasma volume integral is positive. This implies that the drive for vertical instabilities arises from the finite edge $J_\phi$ appearing in the surface integral in $\delta W_F$.

The first goal in our analysis is to convert all volume integrals into surface integrals. This task is accomplished by noting that for $n = 0$ modes the perturbed poloidal fields can be expressed in terms of the perturbed flux in the standard manner. Thus, for each region of interest (i.e. plasma, inner vacuum, and outer vacuum regions) it follows that

$$
\begin{aligned}
\mathbf{B}_{p1} &= \frac{1}{R}\nabla\psi \times \mathbf{e}_\phi \\
B_{p1}^2 &= \frac{(\nabla\psi)^2}{R^2}
\end{aligned}
\tag{2}
$$

with $\psi$ satisfying

$$
\Delta^*\psi = -\left(\mu_0 R^2 p'' + \frac{1}{2}F^{2''}\right)\psi \tag{3}
$$

Clearly, $p'' = F^{2''} = 0$ for the vacuum regions.

Next, use the identity

$$
\nabla\cdot\left(\frac{\psi}{R^2}\nabla\psi\right) = \frac{(\nabla\psi)^2}{R^2} + \frac{\psi}{R^2}\Delta^*\psi = \frac{(\nabla\psi)^2}{R^2} - \left(\mu_0 p'' + \frac{1}{2R^2}F^{2''}\right)\psi^2 \tag{4}
$$

The divergence theorem now allows us to convert all volume integrals into surface integrals, making use of the differential surface element relation $dS = Rd\phi dl = 2\pi R dl$, where $dl$ is the differential poloidal arc length,

$$
L = \frac{\pi}{\mu_0}\int_{S_P}\left[\frac{\psi}{R}\mathbf{n}\cdot\nabla(\psi-\hat{\psi}) + \frac{\mu_0 J_\phi}{RB_p}\psi^2\right]_{S_P} dl + \frac{\pi}{\mu_0}\int_{S_W}\left[\frac{\hat{\psi}}{R}\mathbf{n}\cdot\nabla(\hat{\psi}-\hat{\hat{\psi}}) + \alpha\frac{\hat{\psi}^2}{R}\right]_{S_W} d\hat{l} \tag{5}
$$

Here $dl$ and $d\hat{l}$ are the differential arc lengths along the plasma and wall surfaces respectively. Note that the required continuity of the perturbed fluxes across both the plasma and wall interfaces

$$
\begin{aligned}
\hat{\psi}(S_P) &= \psi(S_P) \\
\hat{\hat{\psi}}(S_W) &= \hat{\psi}(S_W)
\end{aligned}
\tag{6}
$$

has been used to simplify Eq. (5). We point out that equation (5) is valid for arbitrary profiles. The simplification associated with Solov'ev profiles occurs later in the analysis.

## 3. Fourier analysis

The task now is to evaluate $L$ by substituting Fourier series with unknown coefficients for each of the dependent variables. Ultimately the desired relation between elongation and aspect ratio is obtained by standard variational techniques; that is, we set $\delta L = 0$ by varying the Fourier coefficients while simultaneously satisfying the constraint $L = 0$ by iterating $\kappa$ and $\delta$ .

The task of setting $\delta L = 0$ separates into two parts. In the first part, Fourier series are introduced for both the fluxes and their normal derivatives. In the second part, constraint relations between the coefficients in the fluxes and their normal derivatives are obtained by means of Green's theorem. In this section we focus on the first part of the calculation.

The analysis begins by introducing a simple scaling transformation of actual poloidal arc length into an arc length angle. Specifically we write

$$\begin{aligned} l &= \frac{L_P}{2\pi}\chi \\ \hat{l} &= \frac{L_W}{2\pi}\hat{\chi} \end{aligned} \tag{7}$$

Here $L_P, L_W$ are the circumferences of the plasma and wall surfaces respectively. This transformation is convenient because $0 \le \chi \le 2\pi$ and $0 \le \hat{\chi} \le 2\pi$ making it easy to impose poloidal periodicity. The angles $\chi, \hat{\chi}$ are easily determined numerically once the surface coordinates have been specified.

Next, we introduce Fourier series for each of the basic unknowns. For vertical instabilities where $\mathbf{n}\cdot\boldsymbol{\xi}$ has even $Z$ symmetry it follows that the fluxes should be expanded in sine series,

$$\begin{aligned} \psi(S_P) &= \left(\frac{R}{R_0}\right)^{1/2} \sum_1^\infty \psi_m \sin m\chi \\ \hat{\psi}(S_W) &= \left(\frac{R}{R_0}\right)^{1/2} \sum_1^\infty \hat{\psi}_m \sin m\hat{\chi} \end{aligned} \tag{8}$$

As shown shortly the factors in front of the summations simplify the algebra. Each of the unknown normal derivatives is also expanded in a Fourier sine series,

$$\frac{L_P}{2\pi}\mathbf{n}\cdot\nabla\psi(S_P) = 2\left(\frac{R}{R_0}\right)^{1/2}\sum_1^\infty u_n \sin m\chi$$

$$\frac{L_P}{2\pi}\mathbf{n}\cdot\nabla\hat{\psi}(S_P) = 2\left(\frac{R}{R_0}\right)^{1/2}\sum_1^\infty \hat{u}_n \sin m\chi$$

$$\frac{L_W}{2\pi}\mathbf{n}\cdot\nabla\hat{\psi}(S_W) = 2\left(\frac{R}{R_0}\right)^{1/2}\sum_1^\infty \hat{v}_n \sin m\hat{\chi} \tag{9}$$

$$\frac{L_W}{2\pi}\mathbf{n}\cdot\nabla\hat{\hat{\psi}}(S_W) = 2\left(\frac{R}{R_0}\right)^{1/2}\sum_1^\infty \hat{\hat{v}}_n \sin m\hat{\chi}$$

With the required expansions now in hand, we can combine Eqs. (5), (8), and (9) to obtain an expression for the normalized Lagrangian integral $\bar{L} = (\mu_0 R_0 / \pi^2)L$ in terms of the Fourier amplitudes. A short calculation yields

$$\bar{L} = 2\boldsymbol{\psi}^T\cdot(\mathbf{u}-\hat{\mathbf{u}}) + 2\hat{\boldsymbol{\psi}}^T\cdot(\hat{\mathbf{v}}-\hat{\hat{\mathbf{v}}}) + \boldsymbol{\psi}^T\cdot\mathsf{J}\cdot\boldsymbol{\psi} + \gamma\tau_w\hat{\boldsymbol{\psi}}^T\cdot\hat{\boldsymbol{\psi}} \tag{10}$$

where $\boldsymbol{\psi}$ etc. are the vectors of Fourier amplitudes and the elements of the matrix $\mathsf{J}$ can be written as

$$\mathsf{J}_{mn} = \mathsf{J}_{nm} = \frac{1}{\pi}\int_0^{2\pi} d\chi \sin m\chi \sin n\chi \left(\frac{\mu_0 L_P J_\phi}{2\pi B_p}\right)_{S_P} \tag{11}$$

Note the different fonts used for matrices. Also, the precise definition of the wall diffusion time is given by

$$\tau_w = \frac{\mu_0 \sigma d L_W}{2\pi} \tag{12}$$

Equation (10) can be rewritten in the following compact form

$$\bar{L} = \mathbf{x}^T\cdot\mathsf{W}\cdot\mathbf{x} \tag{13}$$

Here, $\mathbf{x}^T = [\boldsymbol{\psi}, \hat{\boldsymbol{\psi}}, \mathbf{u}, \hat{\mathbf{u}}, \hat{\mathbf{v}}, \hat{\hat{\mathbf{v}}}]$ and $\mathsf{W}$ is the symmetric matrix

$$\mathbf{W} = \begin{vmatrix} \mathbf{J} & \mathbf{0} & \mathbf{I} & -\mathbf{I} & \mathbf{0} & \mathbf{0} \\ \mathbf{0} & \gamma\tau_w\mathbf{I} & \mathbf{0} & \mathbf{0} & \mathbf{I} & -\mathbf{I} \\ \mathbf{I} & \mathbf{0} & \mathbf{0} & \mathbf{0} & \mathbf{0} & \mathbf{0} \\ -\mathbf{I} & \mathbf{0} & \mathbf{0} & \mathbf{0} & \mathbf{0} & \mathbf{0} \\ \mathbf{0} & \mathbf{I} & \mathbf{0} & \mathbf{0} & \mathbf{0} & \mathbf{0} \\ \mathbf{0} & -\mathbf{I} & \mathbf{0} & \mathbf{0} & \mathbf{0} & \mathbf{0} \end{vmatrix} \tag{14}$$

Each of the elements in $\mathbf{W}$ is an $M \times M$ matrix with $M$ the number of Fourier amplitudes maintained in the expansions. The total dimensions of $\mathbf{W}$ are thus $6M \times 6M$.

## 4. Application of Green's theorem

The normal derivatives of the fluxes are related to the fluxes themselves through the solution to the vacuum equation $\Delta^*\psi = 0$. (The condition $\Delta^*\psi = 0$ is also true in the plasma region for Solov'ev profiles, and this leads to a much simpler numerical formulation plus savings in computer time). Since the relationships are needed only on the plasma and wall surfaces, a convenient approach is to utilize Green's theorem with the observation point located on either of the surfaces. The procedure is demonstrated below starting with the plasma region. The end results are four linear constraint relations between the various Fourier amplitudes.

- **The plasma region**

In the plasma region the 2-D Green's theorem with the observation point on the plasma surface (i.e. the integration surface) can be obtained from the basic identity

$$\nabla \times \left[ G\nabla \times \left( \frac{\psi}{R}\mathbf{e}_\phi \right) - \psi\nabla \times \left( \frac{G}{R}\mathbf{e}_\phi \right) \right] = G\nabla \times \nabla \times \left( \frac{\psi}{R}\mathbf{e}_\phi \right) - \psi\nabla \times \nabla \times \left( \frac{G}{R}\mathbf{e}_\phi \right) \tag{15}$$

For vacuum fields the flux and 2-D Green's function satisfy

$$\nabla \times \nabla \times \left( \frac{\psi}{R} \mathbf{e}_\phi \right) = 0$$
$$\nabla \times \nabla \times \left( \frac{G}{R} \mathbf{e}_\phi \right) = \delta(R - R')\delta(Z - Z') \tag{16}$$

In these expressions unprimed and primed coordinates refer to the observation and integration points respectively.

The 2-D Green's function is closely related to the flux function for a circular loop of wire. Specifically, the vector potential due to a wire loop, satisfies

$$\nabla \times \nabla \times (A_\phi \mathbf{e}_\phi) = \mu_0 J_\phi \mathbf{e}_\phi = \mu_0 I \delta(R - R')\delta(Z - Z')\mathbf{e}_\phi \tag{17}$$

The solution is

$$RA_\phi = \frac{\mu_0 I}{2\pi}^{1/2} \left( \frac{R'R}{k^2} \right)^{1/2} \left[ (2 - k^2)K - 2E \right]$$
$$k^2 = \frac{4R'R}{(R' + R)^2 + (Z' - Z)^2} \tag{18}$$

Here $K(k), E(k)$ are complete elliptic integrals. Thus, if we set $\mu_0 I = 1$, we see that $RA_\phi = G$,

$$G = \frac{1}{2\pi} \left( \frac{R'R}{k^2} \right)^{1/2} \left[ (2 - k^2)K - 2E \right]$$
$$k^2 = \frac{4R'R}{(R' + R)^2 + (Z' - Z)^2} \tag{19}$$

Also needed in the analysis is the normal derivative (in integration coordinates) of the Green's function evaluated on the plasma surface. A short calculation yields

$$\frac{L_P}{2\pi}(\mathbf{n}' \cdot \nabla' G) = \frac{1}{2}\frac{\dot{Z}'}{R'}(G - G^\dagger) + \frac{\dot{Z}'(R' - R) - \dot{R}'(Z' - Z)}{(R' - R)^2 + (Z' - Z)^2} G^\dagger$$
$$G^\dagger = \frac{1}{2\pi} \left( \frac{R'R}{k^2} \right)^{1/2} \left[ 2(1 - k^2)K - (2 - k^2)E \right] \tag{20}$$

Note that $\dot{Z}', \dot{R}'$ denote $dZ(\chi')/d\chi'$ and $dR(\chi')/d\chi'$ indicating that we have switched integration variables from $l'$ to $\chi'$

The next step is to apply Stokes theorem to Eq. (15) with the observation point on the plasma surface

$$\frac{1}{2}\psi = \int_{L_P}\left[\frac{\psi'}{R'}\nabla' G\times \mathbf{e}_\phi - \frac{G}{R'}\nabla'\psi'\times\mathbf{e}_\phi\right]\cdot d\mathbf{l}' \tag{21}$$

In this expression we need to be a little careful about the signs. The main point is that Stoke's theorem requires $d\mathbf{l}'$ to rotate in a right handed sense. Now, in the usual $R,\phi,Z$ coordinate system this implies that $d\mathbf{l}'$ rotate in the clockwise direction. However, it is convenient and familiar to have $\chi'$ rotate in the counter clockwise direction. Thus, if we define a unit tangential vector $\mathbf{t}'$ pointing in the counter clockwise direction it then follows that

$$\begin{aligned} d\mathbf{l}' &= -\mathbf{t}'dl' = -\frac{L_P}{2\pi}\mathbf{t}'d\chi' \\ \mathbf{t}' &= \frac{\dot{R}'\mathbf{e}_R + \dot{Z}'\mathbf{e}_Z}{(\dot{R}'^2+\dot{Z}'^2)^{1/2}} \\ \mathbf{n}' &= \mathbf{e}_\phi\times\mathbf{t}' = \frac{\dot{Z}'\mathbf{e}_R - \dot{R}'\mathbf{e}_Z}{(\dot{R}'^2+\dot{Z}'^2)^{1/2}} \end{aligned} \tag{22}$$

Here, $\mathbf{n}'$ is the outward pointing unit normal vector. With this sign convention Eq. (21) reduces to

$$\frac{1}{2}\psi = \int_{L_P}\left[\frac{G}{R'}\mathbf{n}'\cdot\nabla'\psi' - \frac{\psi'}{R'}\mathbf{n}'\cdot\nabla' G\right]_{l,l'} dl' \tag{23}$$

A similar expression holds for the wall surface.

The calculation continues by substituting the Fourier expansions into Eq. (23) and then carrying out a Fourier analysis. A straightforward calculation leads to

$$\psi_m + \sum_n A_{mn}\psi_n - \sum_n B_{mn}u_n = 0 \qquad \rightarrow \qquad (\mathbf{I}+\mathbf{A}_{11})\cdot\boldsymbol{\psi} - \mathbf{B}_{11}\cdot\mathbf{u} = 0 \tag{24}$$

where the matrix elements are given by

$$A_{mn} = \frac{2}{\pi}\int d\chi' d\chi \sin n\chi' \sin m\chi \left[\frac{L_P}{2\pi}\frac{\mathbf{n}'\cdot\nabla' G}{(R'R)^{1/2}}\right]_{\chi,\chi'}$$

$$B_{mn} = B_{nm} = \frac{4}{\pi}\int d\chi' d\chi \sin n\chi' \sin m\chi \left[\frac{G}{(R'R)^{1/2}}\right]_{\chi,\chi'} \tag{25}$$

For the matrix format the first subscript on $\mathbf{A}_{11}$ denotes the observation point while the second denotes integration point. This holds for all other matrices that follow.

- **The outer vacuum region**

The analysis of the outer vacuum region is very similar to that of the plasma. One simply has to switch surfaces and take into account the opposite sign of the outward surface normal. The basic equation for the outer vacuum region, assuming regularity at infinity, is given by

$$\frac{1}{2}\hat{\psi} = -\int_{L_W}\left[\frac{G}{R'}\mathbf{n}'\cdot\nabla'\hat{\psi}' - \frac{\hat{\psi}'}{R'}\mathbf{n}'\cdot\nabla' G\right]_{\hat{l},\hat{l}'} d\hat{l}' \tag{26}$$

On this surface the Green's function and its normal derivative are given by

$$G = \frac{1}{2\pi}\left(\frac{R'R}{k^2}\right)^{1/2}\left[(2-k^2)K - 2E\right]$$

$$\frac{L_W}{2\pi}(\mathbf{n}'\cdot\nabla' G) = \frac{1}{2}\frac{\dot{Z}'}{R'}(G-G^\dagger) + \frac{\dot{Z}'(R'-R)-\dot{R}'(Z'-Z)}{(R'-R)^2+(Z'-Z)^2}G^\dagger \tag{27}$$

$$G^\dagger = \frac{1}{2\pi}\left(\frac{R'R}{k^2}\right)^{1/2}\left[2(1-k^2)K-(2-k^2)E\right]$$

The expressions are the same as for the plasma region except that $L_P \to L_W$ in the second equation. Fourier analysis then leads to the following relation between Fourier coefficients

$$\hat{\psi}_m - \sum_n \hat{A}_{mn}\hat{\psi}_n + \sum_n \hat{B}_{mn}\hat{\hat{v}}_n = 0 \qquad \rightarrow \qquad (\mathbf{I} - \mathbf{A}_{22})\cdot\hat{\boldsymbol{\psi}} + \mathbf{B}_{22}\cdot\hat{\hat{\mathbf{v}}} = 0$$

$$\hat{A}_{mn} = \frac{2}{\pi}\int d\hat{\chi}'\, d\hat{\chi}\, \sin n\hat{\chi}' \sin m\hat{\chi}\left[\frac{L_W}{2\pi}\frac{\mathbf{n}'\cdot\nabla' G}{(R'R)^{1/2}}\right]_{\hat{\chi},\hat{\chi}'} \tag{28}$$

$$\hat{B}_{mn} = \hat{B}_{nm} = \frac{4}{\pi}\int d\hat{\chi}'\, d\hat{\chi}\, \sin n\hat{\chi}' \sin m\hat{\chi}\left[\frac{G}{(R'R)^{1/2}}\right]_{\hat{\chi},\hat{\chi}'}$$

- **The inner vacuum region**

The inner vacuum region is slightly more complicated to analyze because of the coupling of surface vectors between the plasma and wall surfaces. In this region Green's theorem must be used twice, once with the observation point on the plasma surface and once on the wall surface. The two basic equations are given by

Observation point on the plasma:

$$\frac{1}{2}\psi(l) = -\int_{L_P}\left[\frac{G}{R'}\mathbf{n}'\cdot\nabla'\hat{\psi}' - \frac{\psi'}{R'}\mathbf{n}'\cdot\nabla' G\right]_{l,l'} dl' + \int_{L_W}\left[\frac{G}{R'}\mathbf{n}'\cdot\nabla'\hat{\psi}' - \frac{\hat{\psi}'}{R'}\mathbf{n}'\cdot\nabla' G\right]_{l,\hat{l}'} d\hat{l}' \tag{29}$$

Observation point on the wall:

$$\frac{1}{2}\hat{\psi}(\hat{l}) = -\int_{L_P}\left[\frac{G}{R'}\mathbf{n}'\cdot\nabla'\hat{\psi}' - \frac{\hat{\psi}'}{R'}\mathbf{n}'\cdot\nabla' G\right]_{\hat{l},l'} dl' + \int_{L_W}\left[\frac{G}{R'}\mathbf{n}'\cdot\nabla'\hat{\psi}' - \frac{\hat{\psi}'}{R'}\mathbf{n}'\cdot\nabla' G\right]_{\hat{l},\hat{l}'} d\hat{l}' \tag{30}$$

After carrying out the Fourier analysis, we arrive at two coupled equations for the Fourier amplitudes

$$\begin{aligned} &\psi_m - \sum_n A_{mn}\psi_n + \sum_n B_{mn}\hat{u}_n + \sum_n \tilde{A}_{mn}\hat{\psi}_n - \sum_n \tilde{B}_{mn}\hat{v}_n = 0 \\ &\hat{\psi}_m + \sum_n \hat{A}_{mn}\psi_n - \sum_n \hat{B}_{mn}\hat{u}_n - \sum_n \tilde{\tilde{A}}_{mn}\psi_n + \sum_n \tilde{\tilde{B}}_{mn}\hat{v}_n = 0 \end{aligned} \tag{31}$$

or in matrix form

$$\begin{aligned} &(\mathbf{I} - \mathbf{A}_{11})\cdot\boldsymbol{\psi} + \mathbf{B}_{11}\cdot\hat{\mathbf{u}} + \mathbf{A}_{12}\cdot\hat{\boldsymbol{\psi}} - \mathbf{B}_{12}\cdot\hat{\mathbf{v}} = 0 \\ &(\mathbf{I} + \mathbf{A}_{22})\cdot\hat{\boldsymbol{\psi}} - \mathbf{B}_{22}\cdot\hat{\mathbf{v}} - \mathbf{A}_{21}\cdot\boldsymbol{\psi} + \mathbf{B}_{21}\cdot\hat{\mathbf{u}} = 0 \end{aligned} \tag{32}$$

The newly introduced matrix elements are defined by

$$\tilde{A}_{mn} = \frac{2}{\pi}\int d\hat{\chi}'\, d\chi \sin n\hat{\chi}' \sin m\chi \left[\frac{L_W}{2\pi}\frac{\mathbf{n}'\cdot\nabla' G}{(R'R)^{1/2}}\right]_{\chi,\hat{\chi}'}$$

$$\tilde{\tilde{A}}_{mn} = \frac{2}{\pi}\int d\chi'\, d\hat{\chi} \sin n\chi' \sin m\hat{\chi} \left[\frac{L_P}{2\pi}\frac{\mathbf{n}'\cdot\nabla' G}{(R'R)^{1/2}}\right]_{\hat{\chi},\chi'}$$

$$\tilde{B}_{mn} = \tilde{B}_{nm} = \frac{4}{\pi}\int d\hat{\chi}'\, d\chi \sin n\hat{\chi}' \sin m\chi \left[\frac{G}{(R'R)^{1/2}}\right]_{\chi,\hat{\chi}'} \tag{33}$$

$$\tilde{\tilde{B}}_{mn} = \tilde{\tilde{B}}_{nm} = \frac{4}{\pi}\int d\chi'\, d\hat{\chi} \sin n\chi' \sin m\hat{\chi} \left[\frac{G}{(R'R)^{1/2}}\right]_{\hat{\chi},\chi'}$$

Note that because of the symmetry $G(R,Z,R',Z') = G(R',Z',R,Z)$ it follows that $\tilde{B}_{mn} = \tilde{\tilde{B}}_{mn}$ implying that $\mathbf{B}_{21} = \mathbf{B}_{12}^T$.

The four constraint relations given by Eqs. (24), (28), and (32) can now be written in a compact form as

$$\mathbf{C}^T \cdot \mathbf{x} = \mathbf{0} \tag{34}$$

where $\mathbf{C}^T$ is a $4M \times 6M$ matrix given by

$$\mathbf{C}^T = \begin{vmatrix} \mathbf{I}+\mathbf{A}_{11} & \mathbf{0} & -\mathbf{B}_{11} & \mathbf{0} & \mathbf{0} & \mathbf{0} \\ \mathbf{0} & \mathbf{I}-\mathbf{A}_{22} & \mathbf{0} & \mathbf{0} & \mathbf{0} & \mathbf{B}_{22} \\ \mathbf{I}-\mathbf{A}_{11} & \mathbf{A}_{12} & \mathbf{0} & \mathbf{B}_{11} & -\mathbf{B}_{12} & \mathbf{0} \\ -\mathbf{A}_{21} & \mathbf{I}+\mathbf{A}_{22} & \mathbf{0} & \mathbf{B}_{21} & -\mathbf{B}_{22} & \mathbf{0} \end{vmatrix} \tag{35}$$

## 5. The numerical solution

The numerical solution to the problem under consideration requires finding stationary solutions to $\mathbf{x}^T \cdot \mathbf{W} \cdot \mathbf{x} = 0$ subject to the constraints $\mathbf{C}^T \cdot \mathbf{x} = \mathbf{0}$. A convenient way to proceed mathematically is to recast the Lagrangian formulation in terms of a minimizing principle by introducing a normalization constraint $\mathbf{x}^T \cdot \mathbf{x} = 1$. Standard linear algebra analysis shows that the Lagrangian formulation is equivalent to (see for instance Trefethen and Bau 1997)

$$\lambda = \frac{\mathbf{x}^T \cdot \mathbf{W} \cdot \mathbf{x}}{\mathbf{x}^T \cdot \mathbf{x}} \qquad \mathbf{C}^T \cdot \mathbf{x} = \mathbf{0} \tag{36}$$

The solution procedure requires a determination of the eigenvalues $\lambda_j$ of $\mathbf{W}$ subject to the constraints $\mathbf{C}^T \cdot \mathbf{x} = \mathbf{0}$. The self-consistency requirement $\mathbf{x}^T \cdot \mathbf{W} \cdot \mathbf{x} = 0$ corresponds to finding (by iteration) a set of plasma parameters $\varepsilon, \kappa, \delta, \beta_p, b/a, \gamma\tau_w$ such that the minimum (i.e. most negative) eigenvalue just happens to satisfy $\lambda_{\min} = 0$.

Practically speaking, once we are able to solve the eigenvalue problem subject to constraints, we can then fix $\beta_p, b/a, \gamma\tau_w$, choose a value for $\varepsilon$, and then iterate to find the largest value of $\kappa$ and corresponding $\delta$ for which $\lambda_{\min} = 0$. In this way the desired curve of $\kappa = \kappa(\varepsilon)$ can be generated.

Finding the eigenvalues of a symmetric matrix $\mathbf{W}$ subject to a set of linear constraints $\mathbf{C}^T \cdot \mathbf{x} = \mathbf{0}$ is a well-known problem in linear algebra. A good way to accomplish this task is by means of a $QR$ orthogonal decomposition (Trefethen and Bau 1997) of the constraint matrix $\mathbf{C}$ and the introduction of a new set of orthonormal basis vectors $\mathbf{z}$ in place of $\mathbf{x}$ (Golub and Underwood 1970). The details of the procedure are given in Appendix A. A summary of the required steps, in the proper sequence, is as follows:

a. Compute (for example using MATLAB 2014) the $QR$ decomposition of $\mathbf{C}$,

$$\mathbf{C} = \mathbf{Q}^T \cdot \begin{vmatrix} \mathbf{R} \\ \cdots \\ \mathbf{0} \end{vmatrix} \tag{37}$$

   Here, the properties and dimensions of the matrices are as follows: $\mathbf{R}$ is a $4M \times 4M$ invertible upper triangular matrix, $\mathbf{0}$ is a $2M \times 4M$ null space matrix, and $\mathbf{Q}$ is a $6M \times 6M$ orthonormal matrix satisfying $\mathbf{Q}^T \cdot \mathbf{Q} = \mathbf{I}$. The symbol L appearing here and in Appendix A is used to indicate the separation between block matrices. Hereafter, we assume that $\mathbf{Q}$ and $\mathbf{R}$ are known matrices.

b. Introduce a new set of orthonormal basis vectors $\mathbf{z}$ in place of $\mathbf{x}$,

$$\mathbf{x} = \mathbf{Q}^T \cdot \mathbf{z} = \mathbf{Q}^T \cdot \begin{vmatrix} \mathbf{z}_4 \\ \mathbf{z}_2 \end{vmatrix} \tag{38}$$

where $\mathbf{z}_4$ contains the first $4M$ elements of $\mathbf{z}$ while $\mathbf{z}_2$ contains the remaining $2M$ elements. Both $\mathbf{x}$ and $\mathbf{z}$ contain a total of $6M$ elements. The analysis in Appendix A shows that the constraint relation forces $\mathbf{z}_4 = \mathbf{0}$.

c. Compute the matrix

$$\mathbf{Q}\cdot\mathbf{W}\cdot\mathbf{Q}^T = \begin{vmatrix} \mathbf{W}_{11} & \mathbf{W}_{12} \\ \mathbf{W}_{12}^T & \mathbf{W}_{22} \end{vmatrix} \tag{39}$$

Here, $\mathbf{W}_{11}$ is $4M\times 4M$, $\mathbf{W}_{22}$ is $2M\times 2M$, and $\mathbf{W}_{12}$ is $4M\times 2M$. Actually only $\mathbf{W}_{22}$ is needed.

d. The desired eigenvalues are obtained from the simplified matrix problem

$$\lambda = \frac{\mathbf{z}_2^T\cdot\mathbf{W}_{22}\cdot\mathbf{z}_2}{\mathbf{z}_2^T\cdot\mathbf{z}_2} \tag{40}$$

The resulting eigenvalue problem automatically takes into account the constraints. Also $\mathbf{W}_{22}$ is symmetric. Its dimensions $2M\times 2M$ are much smaller than the original $\mathbf{W}$ whose size is $6M\times 6M$. Finding the eigenvalues of $\mathbf{W}_{22}$ is a standard numerical problem. In this work, we simply accomplish this task by calling the function "eig" in MATLAB.

The numerical problem has now been fully formulated.

**6. The numerical inputs**

The procedure just described has been implemented in a numerical code that is quick, efficient, and accurate. The parameter space of interest is large, consisting of six physically relevant dimensionless quantities: $\varepsilon$, $\kappa$, $\delta$, $\beta_p$, $b/a$, and $\gamma\tau_w$. The strategy for presenting the results in a compact and understandable form is as follows. First, as a preparatory step we discuss the precise definition of the normalized wall radius parameter $b/a$. Our definition is somewhat different from the usual conformal wall parameter $b/a$. It is more physically realistic in that it holds the normalized gap between inner midplane wall and the plasma $\Delta_i/a$ fixed as the wall area gets larger. Second, after reviewing some experimental data from different tokamaks we define reference values for $\beta_p$, $b/a$, and $\gamma\tau_w$. Once the reference case is established, we compute curves of maximum $\kappa$ and corresponding optimized $\delta$ as a function of $\varepsilon$, separately varying $\beta_p$, $b/a$, and $\gamma\tau_w$.

- **Definition of the normalized wall radius $b/a$**

Our wall model has a shape similar to the plasma. It is characterized by three free input parameters: the normalized inner midplane gap $\Delta_i/a$, the normalized outer midplane gap $\Delta_o/a$, and the normalized vertical gap $\Delta_v/a$. These in turn are easily related to the more familiar normalized wall radius $b/a$ and wall elongation $\kappa_w$. The geometry is illustrated in Fig. 1. In the numerical studies two of the three gap parameters, $\Delta_i/a$ and $\Delta_v/a$, are held fixed. Changing the wall radius corresponds to varying only the outer gap; that is, the single parameter $\Delta_o/a$ or equivalently $b/a$. This choice of variation is motivated by experimental observations (McCracken et al. 1997), which show that the impurity influx in divertor tokamaks from the outboard midplane area is substantially greater than from the inboard side. Consequently, in order to achieve better impurity isolation in future experiments it may be necessary to increase the outboard midplane gap $\Delta_o/a$.

The specific shape of our wall is denoted by the coordinates $\hat{R}, \hat{Z}$ and is given by

$$\begin{aligned} \hat{R} &= \hat{R}_0 + b\cos(\tau + \hat{\delta}_0 \sin\tau) \\ \hat{Z} &= \kappa_w b \sin\tau \end{aligned} \tag{41}$$

Note that the average horizontal wall radius $b/a$ and wall elongation $\kappa_w$ are related to the gap widths and plasma elongation by

$$\begin{aligned} \frac{b}{a} &= 1 + \frac{1}{2}\left(\frac{\Delta_i}{a} + \frac{\Delta_o}{a}\right) \\ \kappa_w &= \frac{a}{b}\left(\kappa + \frac{\Delta_v}{a}\right) \end{aligned} \tag{42}$$

The parameters $\hat{R}_0$ and $\hat{\delta}_0$ can also be expressed in terms of the gap widths by utilizing the assumption that the maximum heights of both the wall and plasma occur at the same $R$.

$$\frac{\hat{R}_0}{R_0} = 1 + \frac{1}{2}\left(\frac{\Delta_o}{a} - \frac{\Delta_i}{a}\right)\varepsilon$$

$$\hat{\delta} = \frac{a}{b}\left[\delta + \frac{1}{2}\left(\frac{\Delta_o}{a} - \frac{\Delta_i}{a}\right)\right] \tag{43}$$

Also, for the numerics it is convenient to normalize and parameterize the wall coordinates as follows: $\hat{R} = R_0\hat{X}, \hat{Z} = R_0\hat{Y}$ with

$$\hat{X} = 1 + \left(\frac{b}{a} - 1 - \frac{\Delta_i}{a}\right)\varepsilon + \left(\frac{b}{a}\right)\varepsilon\cos(\tau + \hat{\delta}_0 \sin\tau)$$

$$\hat{Y} = \left(\frac{b}{a}\right)\kappa_w \varepsilon \sin\tau \tag{44}$$

Once the gap widths and plasma geometry are specified, the wall coordinates given by Eq.(44) are completely determined. From this it is then a straightforward task to calculate the angular arc length coordinate $\hat{\chi}$ on the wall surface.

- **The reference case**

The next step is to define a reference case. The goal is to determine a typical set of values for the parameters of interest: $\beta_p, b/a,$ and $\gamma\tau_w$. To accomplish this task we examine the data for several major large tokamak experiments as shown in Table 1: ASDEX Upgrade (AUG) (Ryter et al. 1998), Alcator C-Mod (C-Mod) (Hutchinson et al. 1994), DIII-D (Lazarus et al. 1991), JET (JET Team 1992), NSTX (Sabbagh et al. 2001), and ITER (Aymar et al. 2002).

| Device / Quantity | AUG | C-Mod | DIII-D | JET | NSTX | ITER |
|---|---|---|---|---|---|---|
| Shot | 12145 | 960214039 | 73334 | 49080 | 132913 | --- |
| $\overline{p}$(atm) | 0.38 | 1.02 | 0.53 | 0.42 | 0.23 | 1.73 |
| $a / R_0$<br>$\varepsilon$ | 0.51/1.60<br>0.32 | 0.23/0.67<br>0.34 | 0.61/1.67<br>0.37 | 0.91/2.91<br>0.31 | 0.58/0.86<br>0.67 | 2.00/6.20<br>0.32 |
| $\kappa$ | 1.84 | 1.77 | 2.05 | 1.93 | 2.42 | 1.72 |
| $\delta$ | 0.28 | 0.70 | 0.85 | 0.36 | 0.66 | 0.49 |
| $b_p$ | 1.37 | 0.70 | 0.86 | 0.84 | 1.11 | 0.38 |
| $\Delta_i / a$ | 0.17 | 0.08 | 0.11 | 0.17 | 0.16 | 0.08 |
| $b / a$ | 1.17 | 1.08 | 1.11 | 1.17 | 1.16 | 1.08 |
| $\kappa_w$ | 2.01 | 1.86 | 2.14 | 2.09 | 2.50 | 1.81 |
| $\kappa_w / \kappa$ | 1.09 | 1.05 | 1.05 | 1.08 | 1.03 | 1.06 |
| $\gamma\tau_w$ | 1.38 | 2.04 | 8.47 | 1.86 | 1.56 | 1.22 |
| $l_i$ | 0.43 | 0.38 | 0.35 | 0.39 | 0.35 | 0.39 |

Table 1. Parameters for high performance elongated tokamaks. Here $\overline{p}$ is the volume averaged pressure and $l_i$ is the internal inductance per unit length of the Solov'ev profile calculated by $l_i = 2\int_{V_p} B_p^2 \, d\mathbf{r} \,/\, (\mu_0^2 I_\phi^2 R_0)$ where $I_\phi$ is the total toroidal current

Each set of data corresponds to a high performance (i.e. high pressure) discharge. Observe first that a reasonable value of poloidal beta for the reference case can be chosen as

$$\beta_p = 1 \tag{45}$$

Next, by combining the experimental data with machine drawings we conclude that the measured inner and outer gap widths are about equal (i.e. $\Delta_o = \Delta_i$) and are set to the values listed in the tables. Also listed is the corresponding value of $b / a$ as calculated from Eq. (42). From the table we then assume that the reference values for the gaps and wall radius are given by

$$\frac{\Delta_i}{a} = \frac{\Delta_o}{a} = 0.1 \quad \rightarrow \quad \frac{b}{a} = 1.1 \tag{46}$$

The reference wall elongation is an additional, but not independent, geometric parameter which enters the analysis but is more difficult to estimate. The walls have different shapes and the spacing between plasma and wall is different on top and bottom because of the divertor. Even for a single experiment it is not clear how to relate the wall shapes in the drawings to the simplified up-down symmetric wall parameter $\kappa_w$.

To circumvent this difficulty we assume that for each experiment the vertical gap is three times the measured inner horizontal gap to allow for a larger vertical space to accommodate the divertor: $\Delta_v / a = 3\Delta_i / a$. Thus, for the table, the reference case, and all future numerical studies, the wall elongation is given by

$$\kappa_w = \frac{a}{b}\left(\kappa + 3\frac{\Delta_i}{a}\right) \tag{47}$$

Here $\kappa$ is arbitrary, $\Delta_i / a$ is specified either experimentally or at its reference value, and $b / a$ is obtained from Eq. (42).

Using the data in Table 1 we have carried out numerical calculations to determine the value of $\gamma\tau_w$ that leads to a numerical eigenvalue $\lambda_{\min} = 0$ for each experiment. By construction, this defines the elongation at high performance that the feedback system, characterized by $\gamma\tau_w$, can safely stabilize. By comparing the $\gamma\tau_w$ data from the different experiments, but omitting DIII-D, we deduce that a typical value of feedback parameter is

$$\gamma\tau_w = 1.5 \tag{48}$$

Interestingly, the value of $\gamma\tau_w$ for DIII-D is substantially higher than for the other experiments and the question is "Why?" We suggest that a much stronger feedback system (i.e. a much larger $\gamma\tau_w$) is needed to achieve the high triangularity $\delta = 0.85$ for the listed shot. Furthermore, this stronger feedback is possible in DIII-D since the feedback coils are located inside the TF coils, much closer to the plasma. In future fusion grade experiments this will probably not be possible because of neutron radiation. This is the reason why a low weight is given to DIII-D when estimating a "reference" value for $\gamma\tau_w$. The DIII-D data is discussed in more detail shortly.

Lastly, we note that we could also deduce a reference internal inductance $l_i$ from the data but this parameter cannot be varied much in our model because of the fixed Solov'ev profiles.

Having defined the reference case we now proceed with a series of numerical calculations to shed insight onto the scaling of maximum achievable $\kappa$ versus $\varepsilon$ as a function of experimental parameters, including the feedback system.

## 7. The numerical results

- **The reference case**

To establish a baseline we calculate curves of maximum $\kappa = \kappa(\varepsilon)$ and the corresponding $\delta = \delta(\varepsilon)$ for the reference case. To do this, we set $\beta_p = 1$, $\Delta_i / a = 0.1$, $\Delta_v / a = 0.3$, $\Delta_o / a = 0.1$ (corresponding to $b / a = 1.1$), and $\gamma\tau_w = 1.5$.

The value of $\kappa_w$ is set in accordance with Eq. (47). The desired scaling curves are obtained by choosing a value for $\varepsilon$ and then iterating on $\kappa$ and $\delta$ such that the eigenvalue $\lambda_{\min} = 0$ for each pair of values. The result can then be plotted as a curve of $\kappa$ versus $\delta$ for the given $\varepsilon$ as shown in Fig. 2. We see that there is an optimized value of $\delta$ for which $\kappa$ is a maximum. Even so, the expanded scale indicates that the maximum is relatively flat in the vicinity of the optimum.

The procedure is repeated for a range of $\varepsilon$ thereby generating a curve of maximum $\kappa = \kappa(\varepsilon)$ and corresponding $\delta = \delta(\varepsilon)$ which is illustrated in Fig. 3. Observe, as expected, that the maximum achievable elongation increases as the aspect ratio becomes tighter. Even so, the $\varepsilon$ dependence is not that strong. As $a / R_0$ increases from 0.1 to 0.8 the maximum $\kappa$ increases from 1.89 to 2.88. The optimum triangularity also increases as the aspect ratio gets tighter but in a stronger way. Over the same range of $a / R_0$ the triangularity increases from 0.05 to 0.65. At $a / R_0 = 0.3$ the maximum elongation and optimum triangularity have the values $\kappa = 2.06$ and $\delta = 0.18$.
The values of $\kappa$ in Fig. 3 are in general slightly higher than those listed in Table 1. The experimental values of $\delta$ are substantially higher. Possible contributing effects to this behavior are as follows. The values in Table 1 correspond to high performance as measured by high average pressure. However, high performance is not determined solely by MHD considerations. Transport plays a comparably important role. Although turbulent transport is known to be reduced with increasing triangularity (Lomas et al. 2000), which helps explain the data, it is unfortunate that the current empirical scaling laws for $\tau_E$ do not explicitly include this effect. Consequently, the highest experimental pressure may be achieved by operating at a larger value of $\delta$ than the MHD optimum because of more than compensatory gains in $\tau_E$.

A second important effect is associated with the fact that any given experiment has a fixed wall shape. Thus, the typical way to increase elongation is by shrinking the minor radius of the plasma. The effective increase in wall radius leads to a higher resistive wall growth rate requiring more feedback and the smaller plasma volume leads to reduced performance because of smaller $\tau_E$. Both lead to a reduced $\kappa$.

A final contributing factor is associated with the fact that the equilibrium Solov'ev current profile used in our analysis is somewhat broader than typical experimental profiles. Specifically, whereas the Solov'ev internal inductance is always about $l_i \approx 0.4$, the more peaked experimental profiles have internal inductances that typically lie in the range $l_i \sim 0.5 - 1.0$. This implies that the Solov'ev profile has a higher current density close to the wall than the experimental profiles, and therefore is more strongly affected by wall stabilization. The result is a slightly higher $k$ for the Solov'ev profile.

Based on this discussion we see that the numerical results presented here and below should be viewed in the context of future experimental designs where the wall to plasma radius can remain fixed as the plasma geometry is varied. Even so, if the designs are based primarily on empirical $\tau_E$ scaling, the impact of triangularity will not be accurately taken into account.

Having established and discussed the reference case we now focus on the scaling of maximum elongation with various physical parameters.

- **Scaling with** $\beta_p$

In the first set of studies as well as all that follow we fix $\Delta_i / a = 0.1, \Delta_v / a = 0.3$. The initial studies focus on scaling with $\beta_p$. As such we fix $\Delta_o / a = 0.1$ (which is equivalent to $b / a = 1.1$) and $\gamma\tau_w = 1.5$. The value of $\kappa_w$ is again set in accordance with Eq. (47).

The desired scaling curves are calculated by repeating the procedure described for the reference case but for various values for $\beta_p$. In Fig. 4 a set of curves is generated for four values of $\beta_p = 0, 0.5, 1, 1.5$. An examination of these curves indicates only a weak scaling of $\kappa$ with $\beta_p$. Noticeable differences occur only for tight aspect ratios, $a / R_0 > 0.5$. With regard to triangularity, observe that the optimum $\delta$ increases with increasing $\beta_p$ although the values, even at $\beta_p = 1.5$, are still below the peak performance values given in Table 1 presumably because of the reasons discussed with the reference case.

A possible reason for the larger triangularity as $\beta_p$ increases is as follows. As $\beta_p$ increases, the contribution to the toroidal current density at the outer-midplane $R > R_0$ becomes larger than the current density at the inner-midplane $R < R_0$. Since the outer midplane toroidal curvature is unfavorable, its effect is minimized by reducing the area on the outside of the plasma. This is accomplished by increasing the triangularity. Hence $\delta$ increases with increasing $\beta_p$. However, $\delta$ cannot become too large because of corresponding increase in unfavorable poloidal curvature at the vertical tips of the plasma.

- **Scaling with** $b / a$

In the second set of studies we fix $\beta_p = 1$, $\gamma\tau_w = 1.5$ and vary the wall radius $b / a$. As previously stated we do this by setting $\Delta_i / a = 0.1, \Delta_v / a = 0.3$ and varying the outer gap parameter $\Delta_o / a$. The values of $b / a$ and $\kappa_w$ are then determined from Eq. (42).

Following the procedure described above we compute curves of $\kappa = \kappa(\varepsilon)$ and the corresponding $\delta = \delta(\varepsilon)$ for various $\Delta_o / a$. These curves are illustrated in Fig. 5 for the values $\Delta_o / a = 0.1, 0.3, 0.5$ or equivalently $b / a = 1.1, 1.2, 1.3$. A comparative plot of the geometries for each elongation is shown in Fig. 6.

The numerical results show that, as expected, moving the wall further out leads to a lower maximum elongation. However, the decrease in maximum elongation is smaller than the increase in wall radius. Specifically, for any $\varepsilon$ a change in $b / a = 0.2$ leads to an approximate change in $\kappa \approx 0.1$. Also, the change in triangularity is small, about 0.05 over the whole range of $\varepsilon$ for the same change in $b / a = 0.2$.

The presumable explanation is that even though the outer part of the wall is being moved further away from the plasma the strong resistive wall image currents stay about the same on the inner, top, and bottom of the first wall since these gaps have been held fixed. In other words the effectiveness of the feedback system is not primarily driven by the proximity of the outer wall to the plasma. One might wonder whether larger decreases in maximum $\kappa$ would occur by instead increasing the inner or upper/lower gaps. This turns out to not be the case based on separate numerical studies that we have carried out (but which for brevity are not reported here). The conclusion is that the maximum $\kappa$ depends significantly on the size of the gap but not its location. .

- **Scaling with $\gamma\tau_w$**

The final set of numerical studies examines the scaling with the feedback parameter $\gamma\tau_w$. For these studies we fix the wall gaps to $\Delta_i / a = 0.1, \Delta_v / a = 0.3, \Delta_o / a = 0.1$ and beta poloidal to $\beta_p = 1$. These are the reference values. The values of $b / a$ and $\kappa_w$ are again determined from Eq. (47).

Curves are generated of $\kappa = \kappa(\varepsilon)$ and the corresponding $\delta = \delta(\varepsilon)$ for $\gamma\tau_w = 0, 1, 2, 3$ as shown in Fig. 7. Observe that the curve for $\gamma\tau_w = 0$ represents an experiment without a vertical stability feedback system. It, therefore, approximates the results for earlier natural elongation studies (see for example Hakkarainen et. al. (1990)). The achievable elongations are indeed quite modest, for example $\kappa = 1.17, \delta = 0.17$ for $a / R_0 = 0.3$.

For higher values of $\gamma\tau_w$ we see that increases in the feedback system capabilities lead to substantial increases in the maximum achievable elongation. Again, for $a / R_0 = 0.3$ the maximum $\kappa$ increases from 1.17 to 2.77 as $\gamma\tau_w$ increases from 0 to 3. The optimized triangularity is insensitive to $\gamma\tau_w$ for small to moderate $e$ but decreases appreciably for tight aspect ratios.

A final quite interesting point concerns a different aspect of triangularity as evidenced in the data from DIII-D in Table 1. To illustrate the point we have carried out a series of calculations assuming a starting point with values of $\kappa = 2.37$ and $\delta = 0.20$ at $\varepsilon = 0.35$ from the $\gamma\tau_w = 2$ curve. We then vary $\delta$ holding $\kappa$ and $\varepsilon$ fixed. At each new $\delta$ we re-compute the value of $\gamma\tau_w$ required to make the eigenvalue $\lambda_{\min} = 0$. This results in a curve of $\gamma\tau_w$ versus $\delta$ as shown in Fig. 8. In other words how much must the feedback capability be increased to stabilize a triangularity that is away from its optimum value? We see that the minimum in $\gamma\tau_w$ is relatively flat in the vicinity of $\gamma\tau_w = 2$ but that a large increase is needed for high triangularities. For example, to achieve a triangularity of 0.71 requires a doubling of the feedback capacity to $\gamma\tau_w = 4$ even though the elongation has remained unchanged. Some insight into this strong behavior can be obtained by noting that the ratio of the pressure driven term to the line bending term in the ideal MHD $\delta W_F$ scales as

$$\frac{2\mu_0(\boldsymbol{\xi}_\perp \cdot \nabla p)(\boldsymbol{\xi}_\perp^* \cdot \boldsymbol{\kappa})}{\left|\mathbf{Q}_\perp\right|^2} \sim \frac{\beta_p}{1-\delta^2} \tag{49}$$

The $1-\delta^2$ factor arises from increasing unfavorable poloidal curvature at the top and bottom of the plasma as $\delta$ becomes larger. This leads to increased instability requiring a larger feedback capability which is consistent with the DIII-D data.

## 8. Discussion

We have calculated the scaling of maximum elongation and corresponding optimized triangularity as a function of inverse aspect ratio for various plasma parameters. The scaling trends are as one might have expected:

- In general, the maximum achievable elongation and optimized triangularity increase as the aspect ratio becomes tighter.
- At fixed aspect ratio the maximum elongation $\kappa_{\max}$, is relatively insensitive to $\beta_p$ except for $\varepsilon \to 1$. For tight aspect ratio, $\kappa_{\max}$ decreases. The optimum triangularity monotonically increases with both $\varepsilon$ and $\beta_p$.
- When the outer midplane wall is moved further away from the plasma then $\kappa_{\max}$ decreases although not by that much. There are still strong image currents on the inner, upper and lower walls to keep the stability largely intact. Also, there is a small increase in triangularity.

- There are large gains in $\kappa_{\max}$ as the feedback capability $\gamma\tau_w$ is increased. This is accompanied by a small to modest decrease in triangularity. One interesting feature is that as the triangularity increases away from its optimum value towards $\delta \to 1$ the required $\gamma\tau_w$ for stability increases rapidly because of the corresponding increase in unfavorable poloidal curvature at the upper and lower tips of the plasma.

Overall, the theoretical predictions of $\kappa_{\max}$ are slightly higher than those observed experimentally for the high performance shots in Table 1. The explanation is likely associated with two effects, both of which effectively increase the experimental wall radius, thereby reducing the achievable $\kappa_{\max}$: (1) shrinking the plasma minor radius to increase plasma elongation and (2) more peaked current profiles than in the Solov'ev model.

A second important theoretical prediction concerns the optimized values of $\delta$ which are noticeably smaller than the observations. The suggestion is that high performance, as measured by high pressure, is not solely dependent on MHD stability. Transport plays a comparably important role in maximizing performance. Gains in $\tau_E$ may more than compensate reductions in $\kappa_{\max}$ by operation away from the optimum $\delta$. Unfortunately, the present empirical scaling relations for $\tau_E$ do not explicitly take into account triangularity. This may therefore be an important challenge for the transport community in the future.

**Acknowledgments**

The authors would like to thank Dr. Martin Greenwald (MIT) for many insightful discussions and a substantial amount of help in obtaining and interpreting tokamak elongation data from many experiments. Thanks are also owed to Drs. Jon Menard (PPPL) and Stan Kaye (PPPL) for their help in understanding the NSTX data as well as to Prof. Dennis Whyte (MIT) for providing the motivation for this work and for many insightful conversations. J. P. Lee and A. J. Cerfon were supported by the U.S. Department of Energy, Office of Science, Fusion Energy Sciences under Award Numbers. DE-FG02-86ER53223 and DE-SC0012398. J. P. Freidberg was partially supported by the U.S. Department of Energy, Office of Science, Fusion Energy Sciences under Award Number DE-FG02-91ER54109.

## Appendix A
## Linear Algebra for $n = 0$ Stability

The stability problem can be written in a classic eigenvalue form as follows

$$\lambda = \frac{\mathbf{x}^T \cdot \mathbf{W} \cdot \mathbf{x}}{\mathbf{x}^T \cdot \mathbf{x}} \tag{A.1}$$

Here $\mathbf{W}$ is an $6M \times 6M$ symmetric matrix and $\mathbf{x}$ is a vector of length $6M$. Also included is the $\gamma\tau_w$ term which enters as an $M \times M$ diagonal matrix contribution to $\mathbf{W}$. The mathematical goal is to find the eigenvalues $\lambda_j$ of $\mathbf{W}$ subject to the Green's function constraints:

$$\mathbf{C}^T \cdot \mathbf{x} = 0 \tag{A.2}$$

The matrix $\mathbf{C}$ has $6M$ rows and $4M$ columns (i.e. $\mathbf{C}$ is a $6M \times 4M$ matrix) and has a rank $4M$. The physical goal requires finding the maximum, $\kappa = \kappa(\varepsilon, \beta_p, b/a, \gamma\tau_w)$ and corresponding $\delta = \delta(\varepsilon, \beta_p, b/a, \gamma\tau_w)$, such that the smallest (i.e. most negative) eigenvalue satisfies $\lambda_{\min} = 0$.

Golub and Underwood have proposed an efficient and elegant method to treat this mathematical problem (Golub and Underwood 1970). The idea is to take into account the constraint relation by carrying out a $QR$ orthogonal decomposition of the constraint matrix $\mathbf{C}$. This allows us to exactly factor out the $4M$ zero eigenvalues arising from the constraint relations, leaving us with a $2M \times 2M$ eigenvalue problem. The $QR$ orthogonal decomposition (called with the function "*qr*" in MATLAB) of $\mathbf{C}$ can be written as

$$\mathbf{C} = \mathbf{Q}^T \cdot \begin{vmatrix} \mathbf{R} \\ \cdots \\ \mathbf{0} \end{vmatrix} \tag{A.3}$$

where, as mentioned in the main text, the symbol $\cdots$ is used to represent the separation between matrix blocks. The properties and dimensions of the matrices, using the notation $m = 6M$, $n = 4M$, and $p = m - n = 2M$ are as follows: $\mathbf{R}$ is an $n \times n$ invertible upper triangular matrix, $\mathbf{0}$ is a $p \times n$ null space matrix, and $\mathbf{Q}$ is an $m \times m$ orthonormal matrix satisfying $\mathbf{Q}^T \cdot \mathbf{Q} = \mathbf{I}$. Note that since $\mathbf{Q}$ is a square matrix it follows that $\mathbf{Q}^T = \mathbf{Q}^{-1}$.

The next step in the procedure, assuming that $\mathbf{Q}$ is known, is to introduce a new set of basis vectors $\mathbf{z}$ in place of $\mathbf{x}$ defined by

$$\mathbf{x} = \mathbf{Q}^T \cdot \mathbf{z} = \mathbf{Q}^T \cdot \begin{vmatrix} \mathbf{z}_n \\ \mathbf{z}_p \end{vmatrix} \tag{A.4}$$

Here, $\mathbf{z}_n$ contains the first $n$ elements of $\mathbf{x}$ while $\mathbf{z}_p$ contains the remaining $p$ elements. Clearly both $\mathbf{x}$ and $\mathbf{z}$ each contain $m$ elements. The usefulness of the transformation becomes apparent when rewriting the constraint relation in terms of $\mathbf{z}$,

$$\mathbf{C}^T \cdot \mathbf{x} = \left( \begin{vmatrix} \mathbf{R}^T & \vdots & \mathbf{0} \end{vmatrix} \cdot \mathbf{Q} \right) \cdot \left( \mathbf{Q}^T \cdot \mathbf{z} \right) = \begin{vmatrix} \mathbf{R}^T & \vdots & \mathbf{0} \end{vmatrix} \cdot \left( \mathbf{Q} \cdot \mathbf{Q}^T \right) \cdot \mathbf{z} = \mathbf{0} \tag{A.5}$$

Now, using the orthonormal properties of $\mathbf{Q}$ it follows that

$$\mathbf{Q}^T \cdot \mathbf{Q} = \mathbf{Q}^{-1} \cdot \mathbf{Q} = \mathbf{Q} \cdot \mathbf{Q}^{-1} = \mathbf{Q} \cdot \mathbf{Q}^T = \mathbf{I} \tag{A.6}$$

Equation (A.5) thus reduces to

$$\mathbf{C}^T \cdot \mathbf{x} = \begin{vmatrix} \mathbf{R}^T & \vdots & \mathbf{0} \end{vmatrix} \cdot \begin{vmatrix} \mathbf{z}_n \\ \mathbf{z}_p \end{vmatrix} = \mathbf{0} \tag{A.7}$$

Carrying out the matrix multiplication leads to the simple result

$$\mathbf{R}^T \cdot \mathbf{z}_n = \mathbf{0} \tag{A.8}$$

Since $\mathbf{R}$ is invertible it has an inverse. Therefore, operating on the left of Eq. (A.8) with $(\mathbf{R}^T)^{-1}$ yields

$$\mathbf{z}_n = \mathbf{0} \tag{A.9}$$

The $QR$ decomposition has led to a set of basis vectors in which the constraint relation is satisfied by the simple step of setting first $n$ elements of $\mathbf{z}$ identically to zero.

We can take this result into account by rewriting the basis vector transformation given by Eq. (A.4) as follows

$$\mathbf{x} = \mathbf{Q}^T \cdot \begin{vmatrix} \mathbf{0} \\ \mathbf{z}_p \end{vmatrix} = \mathbf{Q}^T \cdot \hat{\mathbf{I}} \cdot \mathbf{z}$$

$$\hat{\mathbf{I}} = \begin{vmatrix} \mathbf{0} & \mathbf{0} \\ \mathbf{0} & \mathbf{I}_p \end{vmatrix} \tag{A.10}$$

Observe that $\mathbf{I}_p$ is an identity matrix of dimension $p \times p$ which appears only in the lower right hand corner of the total $m \times m$ matrix $\hat{\mathbf{I}}$. This is a convenient way to suppress the appearance of $\mathbf{z}_n$.

The original eigenvalue problem defined by Eqs. (A.1) and (A.2) can now be simplified by eliminating $\mathbf{x}$ in terms of $\mathbf{z}$

$$\lambda = \frac{\mathbf{x}^T \cdot \mathbf{W} \cdot \mathbf{x}}{\mathbf{x}^T \cdot \mathbf{x}} = \frac{\mathbf{z}^T \cdot \hat{\mathbf{I}} \cdot \mathbf{Q} \cdot \mathbf{W} \cdot \mathbf{Q}^T \cdot \hat{\mathbf{I}} \cdot \mathbf{z}}{\mathbf{z}^T \cdot \hat{\mathbf{I}} \cdot \mathbf{Q} \cdot \mathbf{Q}^T \cdot \hat{\mathbf{I}} \cdot \mathbf{z}} \tag{A.11}$$

The critical point to recognize is that the constraint $\mathbf{C}^T \cdot \mathbf{x} = 0$ is automatically satisfied in this representation. That is, introduction of $\hat{\mathbf{I}}$ eliminates the contribution of $\mathbf{z}_n$ and is equivalent to setting $\mathbf{z}_n = \mathbf{0}$ which is the constraint condition expressed in terms of $\mathbf{z}$.

The numerator and denominator in Eq. (A.11) can be greatly simplified. Using the properties of $\mathbf{Q}$ and $\hat{\mathbf{I}}$ we see that the denominator can be written as

$$\mathbf{z}^T \cdot \hat{\mathbf{I}} \cdot \mathbf{Q} \cdot \mathbf{Q}^T \cdot \hat{\mathbf{I}} \cdot \mathbf{z} = \mathbf{z}^T \cdot \hat{\mathbf{I}} \cdot \hat{\mathbf{I}} \cdot \mathbf{z} = \mathbf{z}_p^T \cdot \mathbf{z}_p \tag{A.12}$$

Next, in the numerator write

$$\mathbf{Q} \cdot \mathbf{W} \cdot \mathbf{Q}^T = \begin{vmatrix} \mathbf{W}_{11} & \mathbf{W}_{12} \\ \mathbf{W}_{12}^T & \mathbf{W}_{22} \end{vmatrix} \tag{A.13}$$

where $\mathbf{W}_{11}$ is $n \times n$, $\mathbf{W}_{22}$ is $p \times p$, and $\mathbf{W}_{12}$ is $n \times p$. Since the starting $m \times m$ matrix $\mathbf{Q} \cdot \mathbf{W} \cdot \mathbf{Q}^T$ is symmetric the matrices $\mathbf{W}_{11}$ and $\mathbf{W}_{22}$ must also be symmetric. Using this information we see that the numerator of Eq. (A.11) reduces to

$$\mathbf{z}^T \cdot \hat{\mathbf{I}} \cdot \mathbf{Q} \cdot \mathbf{W} \cdot \mathbf{Q}^T \cdot \hat{\mathbf{I}} \cdot \mathbf{z} = \mathbf{z}^T \cdot \hat{\mathbf{I}} \cdot \begin{vmatrix} \mathbf{W}_{11} & \mathbf{W}_{12} \\ \mathbf{W}_{12}^T & \mathbf{W}_{22} \end{vmatrix} \cdot \hat{\mathbf{I}} \cdot \mathbf{z} = \mathbf{z}_p^T \cdot \mathbf{W}_{22} \cdot \mathbf{z}_p \qquad \text{(A.14)}$$

Of the total matrix $\mathbf{Q} \cdot \mathbf{W} \cdot \mathbf{Q}^T$ only $\mathbf{W}_{22}$ need be extracted.

The original eigenvalue problem including constraints has now been reduced to the desired form

$$\lambda = \frac{\mathbf{z}_p^T \cdot \mathbf{W}_{22} \cdot \mathbf{z}_p}{\mathbf{z}_p^T \cdot \mathbf{z}_p} \qquad \text{(A.15)}$$

It has been reduced from an $m \times m$ to a $p \times p$ eigenvalue problem for the symmetric matrix $\mathbf{W}_{22}$. Once the eigenvectors have been determined the original vector $\mathbf{x}$ is determined by substituting into Eq. (A.10).

## References

Aymar, R., Barabaschi, P., and Shimomura, Y. 2002 The ITER design *Plasma Physics and Controlled Fusion* **44** 519

Becker, G., Lackner, K. 1977 *Plasma Physics and Controlled Nuclear Fusion Research (Proc. 6th Int. Conf. Berchtesgaden, 1976)* **2** 401.

Bernard, L. C., Berger, D., Gruber, R. and Troyon, F. 1978 Axisymmetric MHD stability of elongated tokamaks *Nuclear Fusion* **18** 1331

Freidberg, J. P. Cerfon, A. and Lee, J. P. 2015 Tokamak elongation – how much is too much? I theory *Journal of Plasma Physics*, submitted

Golub, G.H. and Underwood, R. 1970 Stationary Values of the Ratio of Quadratic Forms Subject to Linear Constraints *Zeitschrift für angewandte Mathematik und Physik ZAMP* **21** 318

Hakkarainen, S. P., Betti, R., Freidberg, J. P. and Gormely, R. 1990 Natural elongation and triangularity of tokamak equilibria *Phys. Fluids B* **2** 1565

Hutchinson, I. H., Boivin R., Bombarda F. et. al. 1994 First results from Alcator-C-MOD *Physics of Plasmas* **1** 1511

JET Team 1992 Fusion energy production from a deuterium-tritium plasma in the JET tokamak *Nuclear Fusion* **32** 187

Lazarus, E. A., Chu, M. S., Ferron, J. R. et. al. 1991 Higher beta at higher elongation in the DIII-D tokamak *Phys. Fluids. B* **3** 2220

Laval, G., Pellat, R. 1973 *Controlled Fusion and Plasma Physics (Proc. 6th Europ. Conf. Moscow 1973)* **2** 640.

Lomas P. J. and JET Team 2000 The variation of confinement with elongation and triangularity in ELMy H-modes on JET, *Plasma Physics and Controlled Fusion* **42** B115

MATLAB version R2014b, Natick, Massachusetts, The MathWorks, Inc. 2014

McCracken, G. M., Lipschultz, B., LaBombard B., Goetz, J. A., Granetz, R. S., Jablonski, D., Lisgo, S., Ohkawa, H., Stangeby, P. C., and Terry, J. L., and the Alcator Group 1997 , Impurity screening in Ohmic and high confinement (H-mode) plasmas in the Alcator C-Mod tokamak *Physics of Plasmas* **4** 1681

Ryter, F., Suttrop, W., Brüsehaber, B. et. al. 1998 H-mode power threshold and transition in ASDEX Upgrade *Plasma Physics and Controlled Fusion* **40** 725

Sabbagh, S.A, Kaye, S.M., Menard, J. et. al. 2001 Equilibrium properties of spherical torus plasmas in NSTX *Nuclear Fusion* **41** 1601

Solov'ev, L. S. 1968 The theory of hydromagnetic stability of toroidal plasma configurations *Sov. Phys.- JETP* **26** 400

Trefethen, L. N. and Bau, D. 1997 Numerical linear algebra, Philadelphia: Society for Industrial and Applied Mathematics

Wesson, J. A., Sykes, A. 1975 *Plasma Physics and Controlled Nuclear Fusion Research (Proc. 5th Int. Conf. Tokyo, 1974)* **1** 449

Wesson, J. A. 1975 *Controlled Fusion and Plasma Physics (Proc. 7th Europ. Conf. Lausanne 1975)* **2** 102.

Wesson, J. A. 1978 Hydromagnetic stability of tokamaks *Nuclear Fusion* **18** 87

## Figure Captions

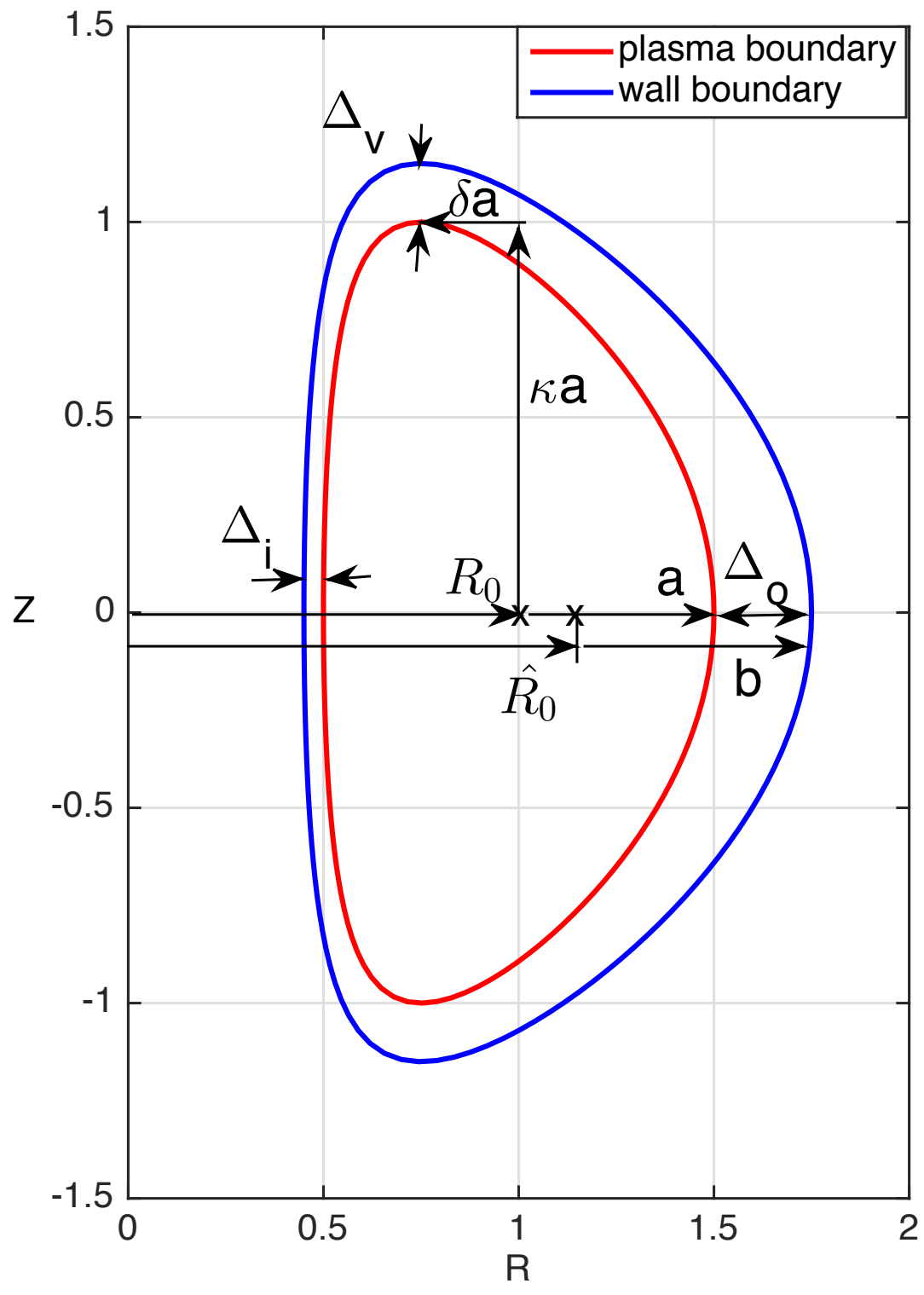

Figure 1 Geometry of the combined plasma – resistive wall system

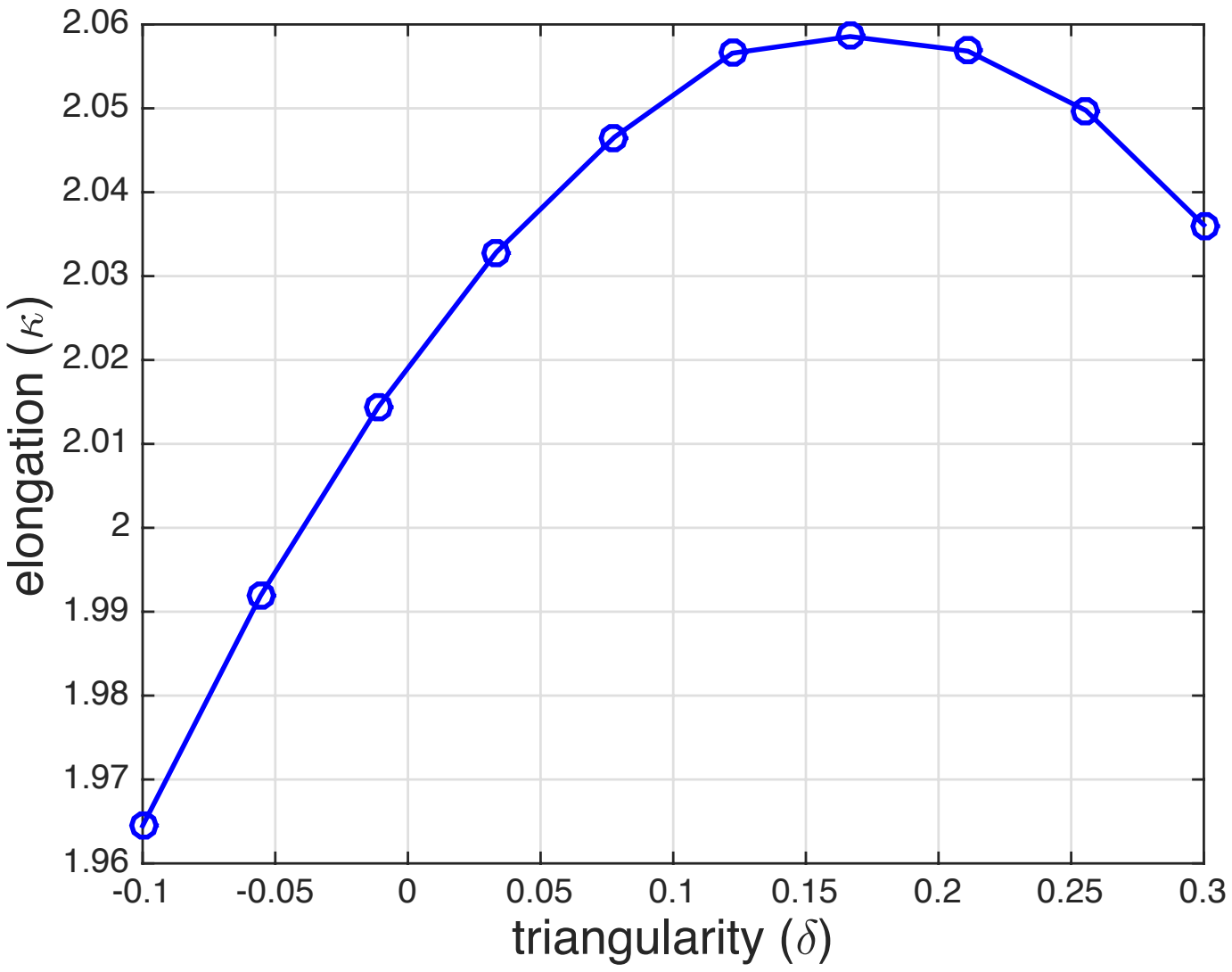

Figure 2 Plot of $\kappa$ versus $\delta$ for $\varepsilon = 0.3$ and the reference values $\beta_p = 1$, $\gamma\tau_w = 1.5$, $\Delta_i / a = \Delta_o / a = \Delta_v / 3a = 0.1$. Observe that there is an optimum $\delta$ at which $\kappa$ is a maximum.

(a) (b)

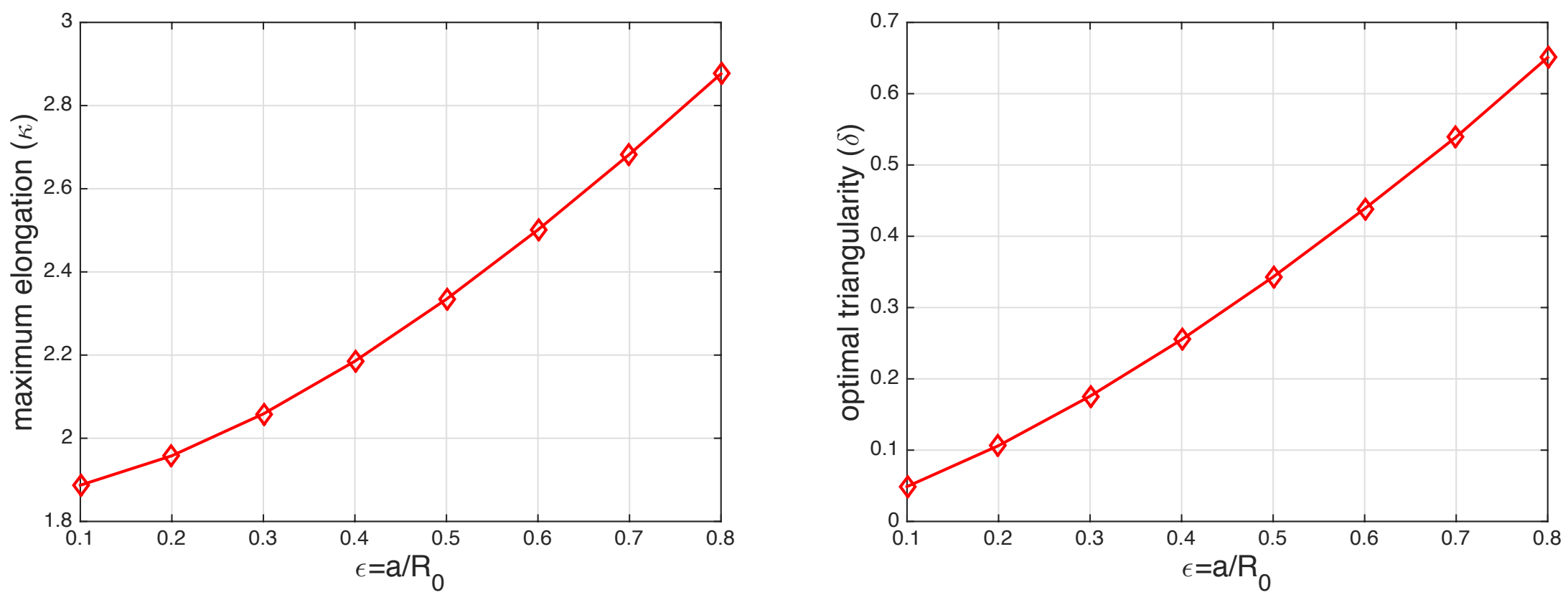

Figure 3 Curves of maximum $\kappa$ (Fig. 3a) and corresponding optimum $\delta$ (Fig. 3b) versus $\varepsilon$ for the reference case $\beta_p = 1$, $\gamma\tau_w = 1.5$, $\Delta_i / a = \Delta_o / a = \Delta_v / 3a = 0.1$.

(a) (b)

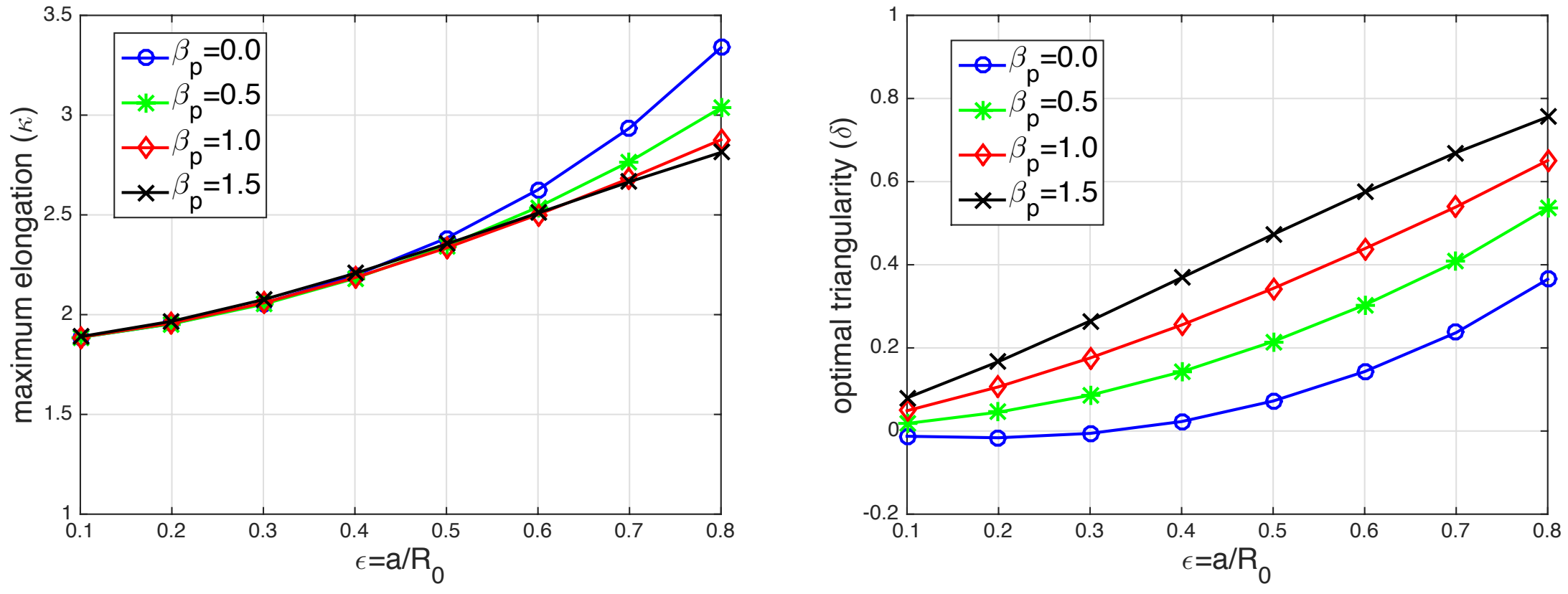

Figure 4 Curves of maximum $\kappa$ (Fig. 4a) and corresponding optimum $\delta$ (Fig. 4b) versus $\varepsilon$ for various values of $\beta_p$ at fixed $\gamma\tau_w = 1.5$, $\Delta_i / a = \Delta_o / a = \Delta_v / 3a = 0.1$.

(a) (b)

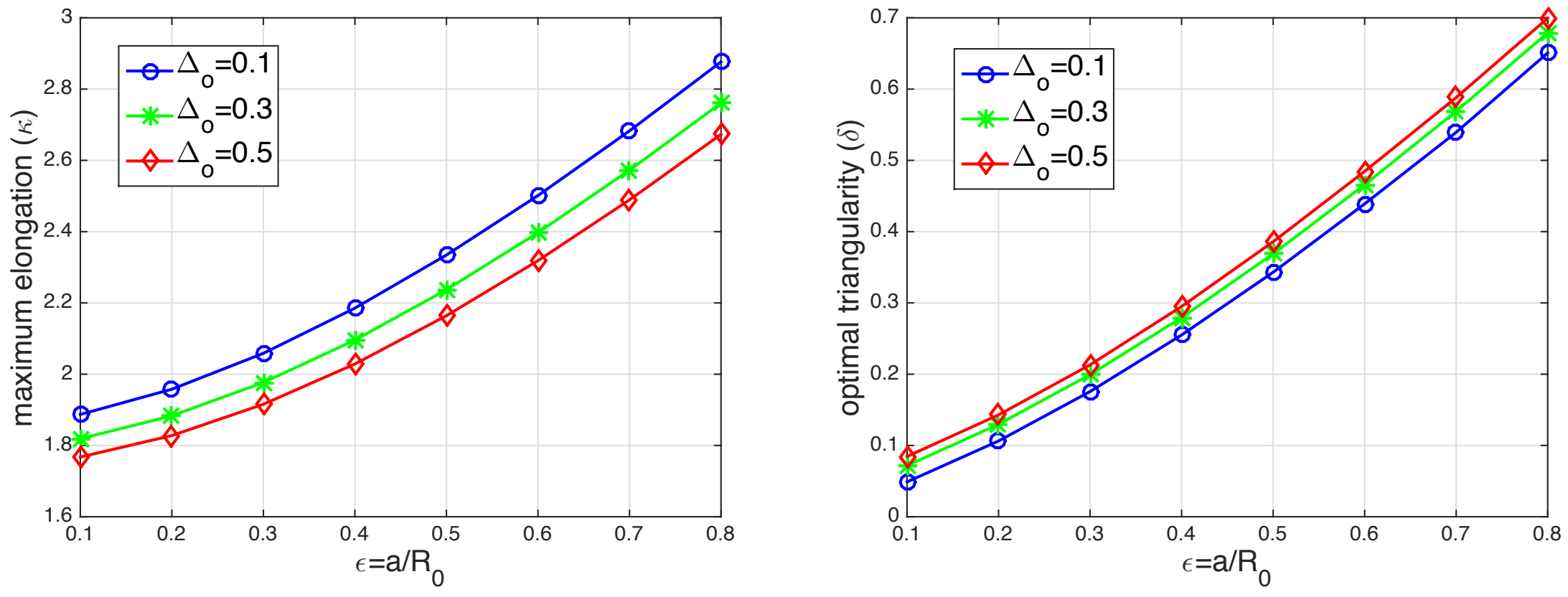

Figure 5 Curves of maximum $\kappa$ (Fig. 5a) and corresponding optimum $\delta$ (Fig. 5b) versus $\varepsilon$ for various values of $\Delta_o / a$ at fixed $\beta_p = 1$, $\gamma\tau_w = 1.5$, $\Delta_i / a = \Delta_v / 3a = 0.1$.

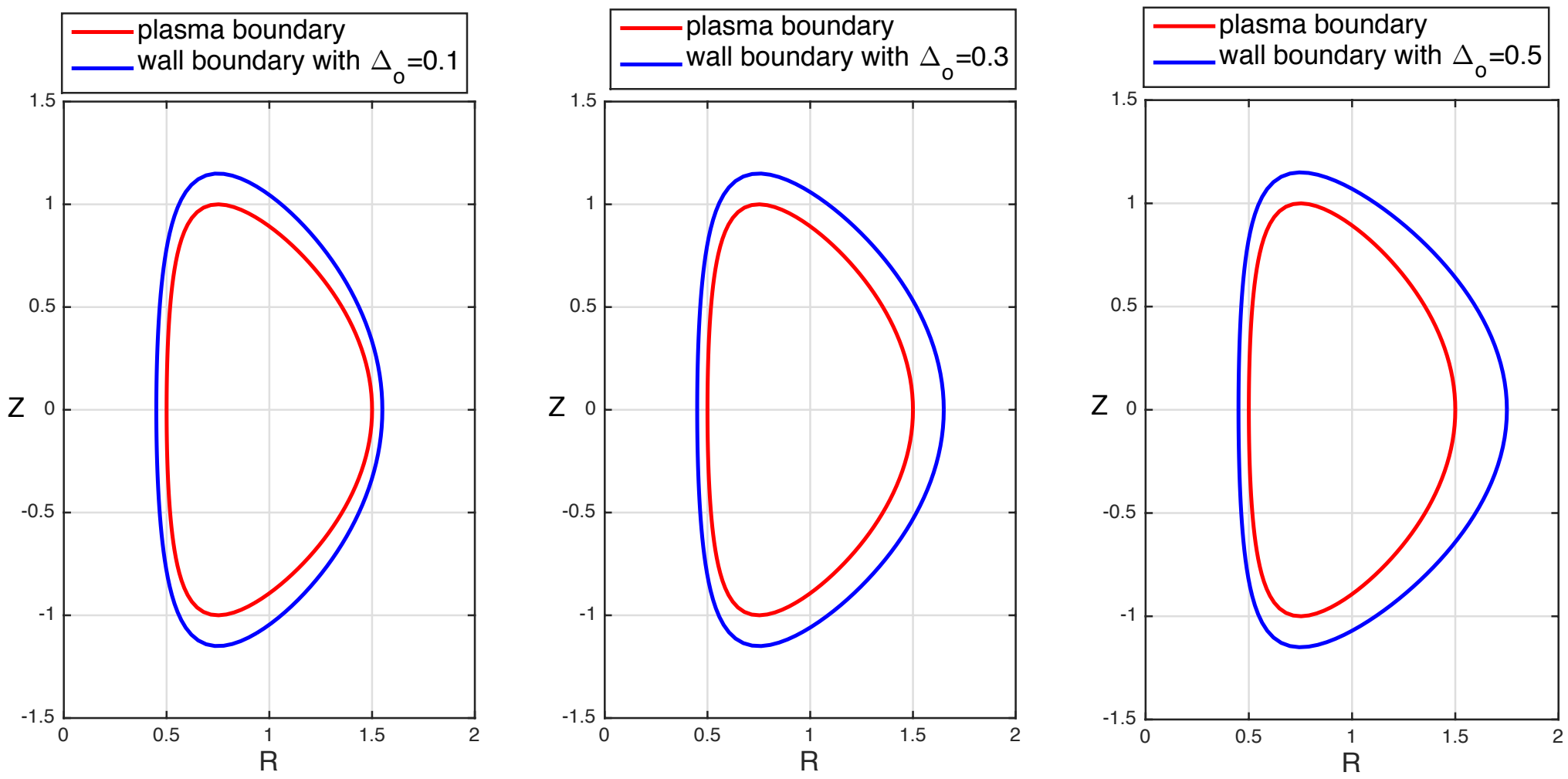

Figure 6 Comparative wall geometries for $\Delta_o / a = 0.1, 0.3, 0.5$ corresponding to $b / a = 1.1, 1.2, 1.3$ at fixed $\Delta_i / a = \Delta_v / 3a = 0.1$.

(a) (b)

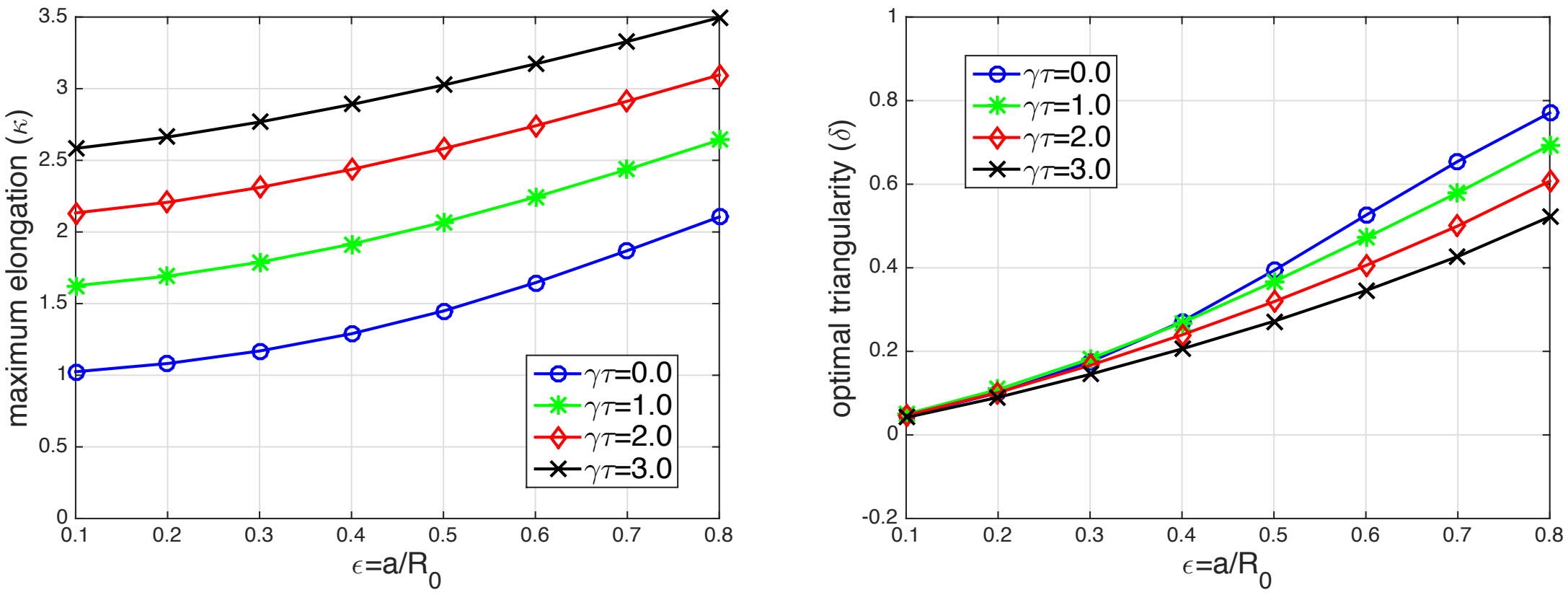

Figure 7 Curves of maximum $\kappa$ (Fig. 7a) and corresponding optimum $\delta$ (Fig. 7b) versus $\varepsilon$ for various values of $\gamma\tau_w$ at fixed, $\beta_p = 1$, $\Delta_i / a = \Delta_o / a = \Delta_v / 3a = 0.1$.

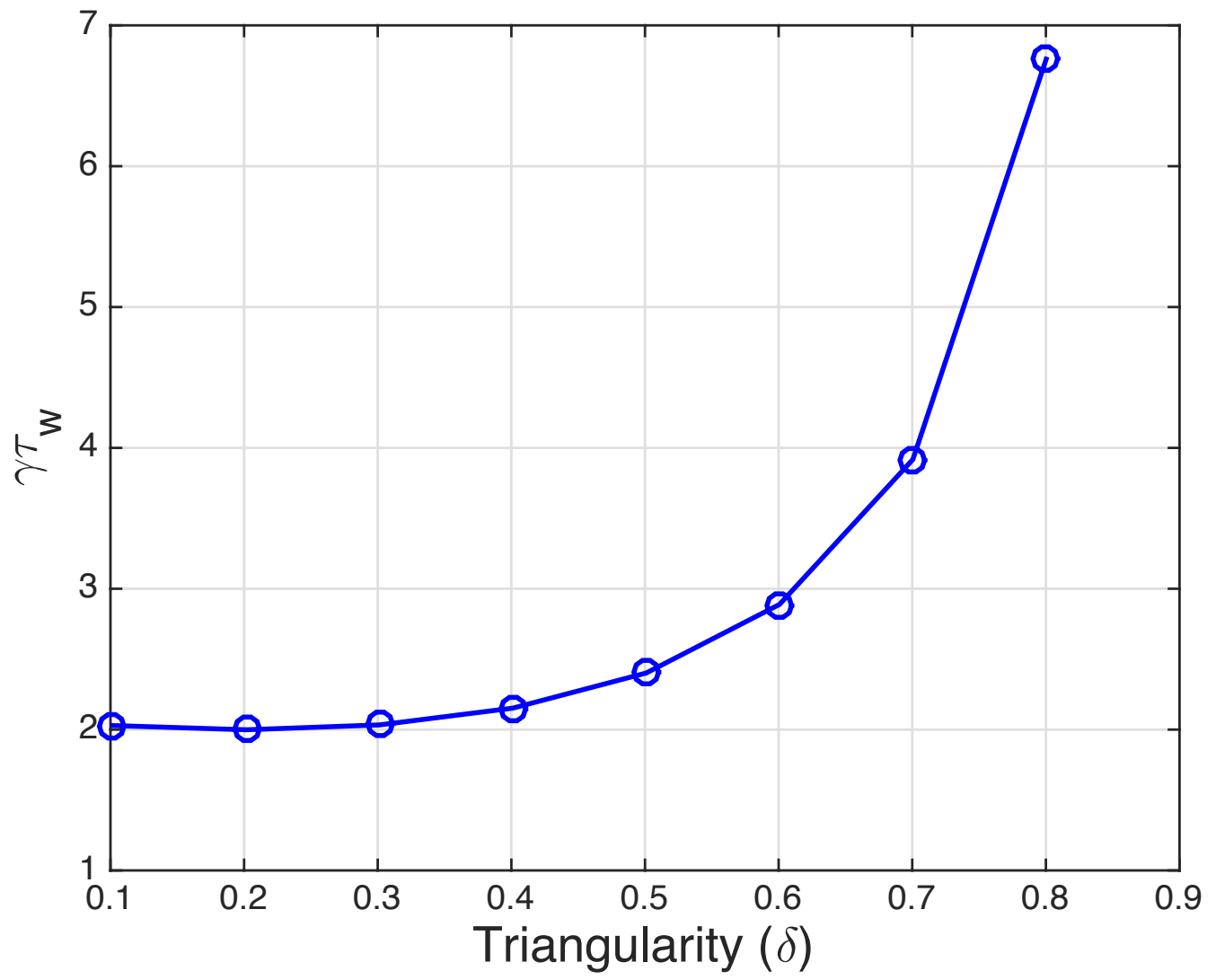

Figure 8 Plot of the required $\gamma\tau_w$ versus $\delta$ at fixed $\varepsilon = 0.35$, $\beta_p = 1$, $\Delta_i / a = \Delta_o / a = \Delta_v / 3a = 0.1$. The minimum in the curve corresponds to $\gamma\tau_w = 2$, $\kappa = 2.37$, and $\delta = 0.20$.